\documentclass[12pt,a4paper]{article}
\pdfoutput=1
\usepackage{graphicx}

\usepackage{jheppub}
\usepackage{cleveref}
\usepackage{subcaption}
\usepackage{lmodern}
\usepackage[normalem]{ulem}

\usepackage{amsmath}
\usepackage{color}

\newcommand{\bd}{\mathrm{bd}}
\newcommand{\be}{\begin{equation}}
\newcommand{\ee}{\end{equation}}
\newcommand{\bea}{\begin{eqnarray}}
\newcommand{\eea}{\end{eqnarray}}
\newcommand{\beal}{\begin{aligned}}
\newcommand{\eeal}{\end{aligned}}

\newcommand{\ep}[1]{ \left( #1 \right) }

\title{Are {\textbf{\textsl{Superentropic}}} black holes superentropic?}
\author[a]{Michael Appels}
\author[b]{Leopoldo Cuspinera}
\author[a,b,c]{Ruth Gregory}
\author[d]{Pavel Krtou\v s}
\author[c]{David Kubiz\v n\'ak}
\affiliation[a]{Centre for Particle Theory, Department of Mathematical Sciences,
Durham University\\
South Road, Durham, DH1 3LE, UK}
\affiliation[b]{Institute for Particle Physics Phenomenology,
Department of Physics,
Durham University\\ South Road, Durham, DH1 3LF, UK}
\affiliation[c]{Perimeter Institute for Theoretical Physics\\
31 Caroline St, Waterloo, Ontario, N2L 2Y5, Canada}
\affiliation[d]{Institute of Theoretical Physics, Charles University\\
V Hole\v sovi\v ck\'ach 2, 180 00 Praha 8, Czech Republic}
\emailAdd{r.a.w.gregory@durham.ac.uk}
\emailAdd{pavel.krtous@utf.mff.cuni.cz}
\emailAdd{dkubiznak@perimeterinstitute.ca}

\abstract{
We study a critical limit in which asymptotically-AdS black holes develop
maximal conical deficits and their horizons become non-compact.
When applied to stationary rotating black holes this limit coincides with
the ``ultraspinning limit'' and yields the \emph{Superentropic} black holes
whose entropy was derived recently and found to exceed the maximal
possible bound imposed by the Reverse Isoperimetric Inequality
\cite{Klemm:2014rda,Hennigar:2014cfa}.
To gain more insight into this peculiar result, we study this limit in the
context of accelerated AdS black holes that have unequal deficits along
the polar axes, hence the maximal deficit need not appear on both
poles simultaneously. Surprisingly, we find that in the presence of acceleration,
the critical limit becomes smooth, and is obtained simply by taking various
upper bounds in the parameter space that we elucidate. The \emph{Critical}
black holes thus obtained have many common features with
\emph{Superentropic} black holes, but are manifestly not superentropic.
This raises a concern as to whether \emph{Superentropic} black holes
actually are superentropic\footnote{We use the upper case {\em Superentropic}
to indicate the specific black hole solution, and lower case superentropic
to indicate the property that entropy violates the Reverse Isoperimetric
Inequality.}. We argue that this may not be so and that the
original conclusion is likely attributed to the degeneracy of the resulting first law.
}
\keywords{black hole thermodynamics, accelerated black holes, reverse isoperimetric inequality}
\preprint{DCPT-19/33}

\begin{document}
\maketitle

\section{Introduction}

Black hole thermodynamics represents a fascinating insight into the interaction
of quantum physics with gravity. Without assigning an entropy to a black hole
\cite{Bekenstein:1973ur}, we would have a violation of the second law of
thermodynamics, widely considered to be one of the most fundamental laws
of nature. Moreover, the discovery of black hole radiation
by Hawking \cite{Hawking:1974sw}, consistent with the
notion of black body radiation, presented definitive proof that black holes
can indeed be assigned quantum properties. As the thermodynamics of
black holes was extended and explored, a natural question was: what is the
black hole equivalent of the pressure/volume term, $PdV$?
Early work \cite{Henneaux:1984ji,Teitelboim:1985dp} proposed that the
cosmological constant $\Lambda$ could fulfil this role, however this was
largely unexplored (though see \cite{Caldarelli:1999xj, Sekiwa:2006qj}) until
the importance of anti-de Sitter (AdS) spacetime came to the fore in the
context of the gauge-gravity duality in string theory. A crucial conceptual
insight was that the `mass', $M$, for the black hole should more properly
be interpreted as enthalpy \cite{Kastor:2009wy}, the pressure with the (negative)
cosmological constant, $P=-\Lambda/(8\pi)$, and
the black hole volume with the corresponding conjugate quantity,
$V=\partial M/\partial P$ \cite{Dolan:2010ha,Dolan:2011xt,Kubiznak:2012wp},
and with this, the subject enjoyed a renaissance,
with many interesting critical phenomena and thermodynamic processes
being explored, see \cite{Kubiznak:2016qmn} for a review.

Within the context of extended black hole thermodynamics there has
been an interesting conjecture --- the {\it Reverse Isoperimetric Inequality}
\cite{Cvetic:2010jb}, which is a statement about the relation between the
thermodynamic volume of the black hole and its entropy, or area.
In mathematics, the {\it Isoperimetric Inequality} states that the surface
area enclosing a given volume is minimised for a spherical surface,
and indeed the area can be unboundedly
large if a suitably deformed or wrinkly surface is chosen. From the
physical perspective of black hole thermodynamics, however, this would
be a disturbing inequality as, if true, the second law would imply that a
black hole would want to be as deformed as possible to maximise its
entropy, thus indicating a classical instability of black holes.
However, in Cvetic et al.~\cite{Cvetic:2010jb},
it was demonstrated that in all
(then) studied black hole solutions, the reverse of this inequality held,
hence the  {\it Reverse Isoperimetric Inequality Conjecture}
(see also \cite{Dolan:2013ft} for the de Sitter version of this conjecture).

Not long after a rather peculiar solution was investigated.
In exploring possible black hole solutions in four-dimensional Fayet--Iliopoulos
gauged supergravities, Gnecchi et al.~\cite{Gnecchi:2013mja} briefly presented a
black hole with a novel horizon topology. The solution emerged as a certain limit
of the Carter--Plebanski metric \cite{Carter, Plebanski} where the metric
function governing the longitudinal angle develops a certain double root.
That it can be interpreted as
the \emph{ultra-spinning} limit of the Kerr-AdS solution, where the rotation
parameter $a$ is taken to be critically large (equal to the AdS radius $\ell$)
was suggested in a letter by Klemm~\cite{Klemm:2014rda}, and the
corresponding limiting procedure was explicitly found in
\cite{Hennigar:2014cfa, Hennigar:2015cja,Hennigar:2015gan}.
The result is a non-compact horizon of finite area, which is roughly
spherical near its equator but becomes hyperbolic near the axis.
The poles are removed from the spacetime and the
horizon topology is that of a sphere with two punctures.

In a series of papers, Hennigar et al.~\cite{Hennigar:2014cfa,
Hennigar:2015cja,Hennigar:2015gan} explored the thermodynamic
implications of having such an extraordinary spacetime. These papers
argued a distinct definition of thermodynamic variables from the standard
Kerr-AdS variables, and intriguingly discovered that the black hole
appeared to be \emph{superentropic}. Specifically, the \emph{reverse
isoperimetric conjecture}~\cite{Cvetic:2010jb,Dolan:2013ft} was found to be
violated by the ultra-spinning black hole, leading the authors to impose
more stringent conditions under which the bound might be valid
-- the {\em Superentropic} black holes
with non-compact horizons had to be excluded from the conjecture.

In this paper, we seek to determine the uniqueness of this latter discovery. A
curious feature of the ultra spinning spacetime is that it is seemingly isolated
from regularly-spinning black holes by any physical process. It is interesting
therefore to ponder whether it truly is a special case, or whether this
violation is present in further extensions of this solution. One way in which
the set of black hole solutions can be extended beyond the usual generalisations
to charged and/or rotating solutions is to consider acceleration.
The solution that describes the accelerated black hole is known as the
C-metric \cite{ehlers1962exact,Kinnersley:1970zw,Plebanski:1976gy,
Griffiths:2005qp,Podolsky:2002nk,Krtous:2005ej, Griffiths:2006tk}.
It is similar in form to Kerr-AdS, but has conical defect(s) along the polar
axes that are different in magnitude, the differential deficit providing a
nett force on the black hole, hence acceleration.

To probe superentropicity in this setting, we will re-visit the ultraspinning
limit of Kerr-AdS black holes discussed above.
We show that the same {\em Superentropic} black holes are obtained by running
a conical deficit through the Kerr-AdS spacetime\footnote{In a physical
picture, one might think about such a conical deficit as being caused by
a cosmic string threading the black hole, though it is not entirely clear
that this is a complete story for rotating black holes
\cite{Gregory:2013xca, Gregory:2014uca}.},
and making this conical deficit maximal, equal to
$2\pi$. This provides an alternative to the ultraspinning limit that is also
applicable to (not necessarily spinning) accelerated black holes.
While the characteristic feature of the ultra-spinning black hole is the
pair of maximal deficits at
each pole, the accelerated solution has by default one deficit greater
than the other, which means that we may only have one such maximal
defect. Further, because the conical defects are present \emph{a priori},
it is possible to maximise one simply by choosing a suitable values of the
mass parameter, independent of whether the black hole is charged or
rotating \cite{Chen:2016rjt, Chen:2016jxv}.
The term ``ultra-spinning'' is therefore no longer appropriate to
designate these special solutions, so we will use the term
\emph{critical} (for lack of an original word) to designate any black hole
solution which exhibits a single (or a pair of) $2\pi$-conical deficit(s).

The thermodynamic properties of black holes in AdS have been known for a
while~\cite{Gibbons:2004ai,Silva:2002jq,Caldarelli:1999xj,Hawking:1998kw,
Kostelecky:1995ei,Hawking:1982dh}, however the thermodynamics of accelerating
black holes have only more recently been elucidated \cite{Appels:2016uha,
Appels:2017xoe,Gregory:2017ogk,Anabalon:2018ydc,Anabalon:2018qfv,
Gregory:2019dtq,Astorino:2016xiy,Astorino:2016ybm}, and in the context of
thermodynamics, we observe these accelerated critical
solutions to behave differently to the original ultra-spinning black hole.
For the ultra-spinning case, the thermodynamic quantities cannot
simply be obtained by taking the $a\to \ell$ limit of the thermodynamic
quantities of Kerr-AdS black holes, as these diverge in the limit $a\to\ell$.
Instead, the thermodynamics of ultra-spinning black holes were
constructed in \cite{Hennigar:2014cfa, Hennigar:2015cja,Hennigar:2015gan}
``afresh'', starting from the {\em Superentropic} metric and applying the standard
procedures, such as the method of conformal completion \cite{Ashtekar:1999jx}.
In this way, a new set of consistent (and finite) thermodynamic quantities,
that are evidently disconnected from those of Kerr-AdS black holes,
were obtained and shown to satisfy the corresponding (degenerate) first
law, and violate the reverse isoperimetric inequality.
In contrast, here we find that when accelerated black holes are critical,
their thermodynamic quantities can be obtained as a smooth (and finite)
limit of the original thermodynamic quantities for the accelerated
black holes \cite{Appels:2017xoe,Appels:2016uha}.
It then follows that the reverse isoperimetric conjecture, shown to be valid
for the accelerated black holes \cite{Gregory:2019dtq}, remains true also
for the critical black holes.

The discrepancy of the two results is astonishing. We attribute it to the
two interrelated basic facts. First, a crucial step towards establishing
thermodynamics of rotating black holes is to correctly identify a possible
rotation at infinity $\Omega_\infty$ \cite{Caldarelli:1999xj, Gibbons:2004ai}.
This affects the conjugate quantity to the angular momentum, and in its
turn also modifies the thermodynamic mass. In the case of ultra-spinning
black holes it is very difficult to identify $\Omega_\infty$ as this formally
diverges; a particular choice of $\Omega_\infty=0$ was made in
\cite{Hennigar:2014cfa, Hennigar:2015cja,Hennigar:2015gan}.
Second, due to the ultraspinning limit $a\to \ell$, the (would be) mass
or enthalpy ${M_S}$ is directly dependent on the (would be) angular
momentum charge ${J_S}$, the two obeying the `chirality condition'
${J_S}={M_S}\ell$ that clearly interrelates $M_S, J_S$, and
$P$ \cite{Klemm:2014rda}. The corresponding first law is thus degenerate and
the thermodynamic quantities in it are no longer uniquely defined.
To illustrate this point, in Sec.~\ref{sec:comp} we construct a different
set of thermodynamic quantities for ultraspinning black holes, that are
also consistent and can be derived from the standard methods (with
different $\Omega_\infty$), but do not violate the reverse isoperimetric
inequality. Together with the results on the thermodynamics of the
critical black holes this raises an interesting question: are the original
{\em Superentropic} black holes truly superentropic?

In the next section we review the accelerating black hole geometry,
focussing on the slowly accelerating black hole \cite{Podolsky:2002nk},
discuss the corresponding admissible parameter space, conical deficits,
and thermodynamics.  In Sec.~\ref{sec:critTD}, we construct the critical
black holes and confirm that they obey the reverse isoperimetric
inequality. In Sec.~\ref{sec:comp} we compare the obtained accelerating results
to those of ultra-spinning black holes; a novel `derivation' of the ultra-spinning
thermodynamics is presented and subjected to critical comments.
We summarise in Sec.~\ref{sec:Concl}, and discuss the example
of superentropic thermodynamics in the BTZ black hole. 

\section{Accelerated black holes} \label{sec:accbh}

\subsection{The generalized C-metric}

The geometry of an accelerating black hole is given by the ``C-metric''
(so called because of a classification scheme of axisymmetric metrics
\cite{ehlers1962exact}) that describes a local black hole type of horizon,
distorted by conical deficits that provide the accelerating force
acting on the black hole \cite{Kinnersley:1970zw,Plebanski:1976gy}.
In anti de Sitter (AdS) spacetime, where $\ell = \sqrt{|\Lambda|/3}$ is
the AdS lengthscale, the metric can be written in the following form
\cite{Hong:2003gx,Hong:2004dm,Chen:2015vma}:
\be
\beal
ds^2 = & \frac{f(r)}{\Sigma H^2}\Big[
\frac{dt}{\alpha}-a\,\sin^2\theta \frac{d\phi}{K} \Big]^2
- \frac{\Sigma dr^2}{f(r)H^2} \; - \frac{\Sigma r^2}{g(\theta)H^2}d\theta^2\\
&- \frac{g(\theta)\sin^2\theta}{\Sigma r^2 H^2} \Big[\frac{adt}{\alpha}-(r^2+a^2)
\frac{d\phi}{K}\Big]^2\,,
\eeal
\label{eq:metric}
\ee
where the metric functions are
\be
\beal
f(r) &=(1-A^2r^2)\left[1-\frac{2m}{r}+\frac{a^2+e^2}{r^2}\right]
+\frac{r^2+a^2}{\ell^2}\,,\\
g(\theta) &=1+2mA\cos\theta+ (\Xi-1)\cos^2\theta\,,\\
\Sigma &=1+\frac{a^2}{r^2}\cos^2\theta\,, \qquad
H=1+Ar\cos\theta \,,\\
\Xi &= 1 + e^2 A^2 - \frac{a^2}{\ell^2} (1-A^2 \ell^2)\,,
\eeal
\ee
and the electromagnetic potential is given by
\be
B=-\frac{e}{\Sigma r}\Big[{dt\over \alpha}-a\sin^2\theta \frac{d\phi}{K}\Big]
+ \Phi_{t} dt\,, \qquad \Phi_{t} = \frac{e r_+}{\alpha(a^2 + r_+^2)}\,.
\label{bpotential}
\ee
The parameters $a$, $e$, $m$, and $A\geqslant 0$ are related to the
angular momentum, charge, mass and acceleration of the black hole, respectively.
It is worth commenting on a few aspects of this geometry before turning
to the features we will be exploring in the next section.

Note the presence of the parameter $K$ associated with the $\phi$
coordinate. In \eqref{eq:metric}, the range of the angular parameter
$\phi$ is taken to be $2\pi$, thus the parameter $K$ will encode in part
the conical deficits along each axis. Next, note that the time coordinate
has been rescaled by $\alpha$. It might seem therefore that a new
parameter has been introduced, however, because the time coordinate
is non-compact, the rescaling by $\alpha$ represents a gauge degree
of freedom: time is usually chosen relative to an asymptotic observer,
which for the accelerating black hole is not entirely straightforward to
define. In \cite{Anabalon:2018qfv}, using holographic renormalization, this
was found to be
\be
\alpha=\frac{\sqrt{(\Xi+a^2/\ell^2)(1-A^2 \ell^2\Xi)}}{1+a^2A^2}\,.
\label{alphadef}
\ee

The conformal factor, $H$, sets the location of the boundary at
$r_\bd=-1/A\cos\theta$, that lies ``beyond infinity'' for $\theta<\pi/2$.
The coordinates in \eqref{eq:metric} therefore do not cover the full
spacetime (though can easily be extended -- see subsection \ref{sec:parameters}),
but are nonetheless useful coordinates as they intuitively extend the
familiar Kerr metric to include acceleration. Finally, note that usually
a uniformly accelerating observer has an acceleration horizon,
however, if $A\ell<{\cal O}(1)$ (again see \S \ref{sec:parameters}),
the function $f$ is positive outside the black
hole event horizon, suggesting that there is no acceleration horizon
and the black hole is simply suspended in AdS at a finite displacement
from the centre. This is known as a \emph{slowly accelerating} black hole
\cite{Podolsky:2002nk}, and will be the focus of our study, although the
actual bound on $A\ell$ is slightly modified to account for the lack
of an acceleration horizon beyond $r=\infty$ as we describe below.
We now turn to this, and other parametric restrictions before discussing
the conical deficit structure and the critical limit.

\subsection{Coordinate ranges and parametric restrictions}
\label{sec:parameters}

To explore what restrictions might apply to the parameters in this metric,
we must translate the physical requirements for the slowly rotating black
hole into statements about the functions $f(r)$ and $g(\theta)$ that then
give constraints on the parameters in the metric. That we are dealing
with a black hole means that we have a zero for $f(r)$ that corresponds
to $2m$ in the limit that $\ell\to\infty$, $e,a,A\to0$, and lies entirely
inside the AdS bulk. That the black hole
lacks an acceleration horizon means that there is no other relevant
zero of $f$. Finally, that $\theta$ corresponds to the angular coordinate
on the (deformed) 2-sphere requires that $g(\theta)\geq0$ on $[0,\pi]$.

The constraint that there is a black hole horizon corresponds to the
existence of an $r_+$ such that $f(r_+)=0$, with $f'(r_+)\geq0$, and
that this horizon lies fully within the spacetime. The former requirement
is relevant in the case of a charged or rotating black hole, and
corresponds to the black hole being sub-extremal, or extremal
if $f'(r_+)=0$. The latter requirement translates to $Ar_+<1$; as
otherwise it would be possible for $1/Ar_+=-\cos\theta_+$ for some
$\theta_+$, hence the event horizon would reach the boundary.

To explore these constraints, for convenience set the scale of the
dimensionful parameters using the acceleration:
\be
\tilde{r} = Ar\,,\quad
\tilde{m} = Am\,,\quad
\tilde{e} = Ae\,,\quad
\tilde{a} = Aa\,,\quad
\tilde{\ell} = A\ell\,.
\ee
We can now solve the extremality constraint $f(r_+) = f'(r_+)=0$ leading
to constraints on the mass and cosmological constant (i.e.\ $\ell$) expressed in
terms of the charge and angular momentum (or vice versa). These can
conveniently be parametrised in terms of horizon radius:
\be
\tilde{m} = \frac{(\tilde{r}_+^2+\tilde{a}^2)^2 +
\tilde{e}^2(\tilde{a}^2 - \tilde{r}_+^4+ 2 \tilde{r}_+^2)}
{\tilde{r}_+\left(\tilde{a}^2(1+\tilde{r}_+^2) + \tilde{r}_+^2(2-\tilde{r}_+^2)\right)}\,,\quad
\tilde{\ell}^2 = \frac{ \tilde{r}_+^2
( \tilde{r}_+^2 -  \tilde{a}^2  \tilde{r}_+^2 - 3  \tilde{r}^2_+ -  \tilde{a}^2)}
{ (1- \tilde{r}_+^2)^2(\tilde{r}_+^2 -  \tilde{a}^2 -  \tilde{e}^2)}\,.
\ee

In order to explore the constraint from slow acceleration, note that outside
the black hole horizon $f(r)$ is positive, but
while $r$ is a familiar coordinate for describing the properties
of the black hole, it does not cover the full spacetime, instead
$y=-1/Ar$, running from $-1/Ar_+$ on the horizon to $\cos\theta$ on the
boundary proves to be a better coordinate. The region of spacetime
beyond $r=\infty$ is now covered by positive values of $y$, and the lack
of an acceleration horizon in this region corresponds to $F(y)>0$, where
\be
F(y) =  \tilde{\ell}^2 y^2 f(-1/Ay)
= 1 + \tilde{a}^2 y^4 -  \tilde{\ell}^2 (1-y^2)
\left (1 + 2  \tilde{m} y + ( \Xi - 1) y^2 \right)\,.
\ee
$F$ has a minimum on $[0,1]$, so the borderline case as the acceleration
horizon forms is $F(y_0) = F'(y_0)=0$, giving
\be
\tilde{m} = y_0
\frac{(1+\tilde{a}^2 y_0^2)^2 - \tilde{e}^2 (1 - 2y_0^2 - \tilde{a}^2 y_0^4)}
{1 - y_0^2 \left(3+ \tilde{a}^2 (1+y_0^2) \right)}\,,\quad
\tilde{\ell}^2 =  \frac{1-3y_0^2 - \tilde{a}^2 y_0^2 (1+y_0^2)}
{ (1-y_0^2)^2 \left(1-y_0^2(\tilde{a}^2 + \tilde{e}^2)\right)}\,.
\ee

Finally, the constraint that $g(\theta)\geq0$ on $[0,\pi]$, i.e.
\be\label{eq:gx}
1+ 2\tilde m x + \left(\Xi-1\right) x^2
\geq0\qquad \text{for} \quad x \in [-1,1]
\ee
translates to
\be
\tilde{m} \leq
\begin{cases}
\Xi/2 & \Xi\leq 2 \\
\sqrt{\Xi-1} & \Xi>2\,.
\end{cases}
\label{eq:conditions}
\ee
However, the requirement that $Ar_+<1$ implies that the term in $f(r_+)$
inside square brackets is negative:
\be
\tilde{r}_+^2 - 2\tilde{m} \tilde{r}_+ + \tilde{e}^2 + \tilde{a}^2 < 0\,.
\ee
Clearly this quadratic must have real roots, and this
in turn requires that its discriminant be positive:
\be
\tilde{m}^2 > \tilde{e}^2 + \tilde{a}^2 = \Xi-1+\frac{\tilde{a}^2}{\tilde{\ell}^2}
\geq \Xi-1\,,
\ee
in clear contradiction with \eqref{eq:conditions} for $\Xi\geq 2$.
Thus, the constraints arising from the angular coordinate
require
\be
\Xi < 2 \qquad \text{and} \qquad \tilde{m} \leq \Xi/2\,.
\ee

To sum up: the constraint from $g(\theta)$ gives an upper bound on $\tilde{m}$,
the constraint from extremality gives a lower bound on $\tilde{m}$, and the
constraint from slow acceleration gives an upper bound on $\tilde{\ell}$, that is
$\tilde{m}$-dependent.

\subsection{The conical defect}

The presence of a conical deficit in the spacetime is parametrised (in part)
by the parameter $K$. Whether or not there is acceleration, if $K\neq1$,
the metric will not be flat along at least one of the axes. To see this,
expand the angular part of the metric in \eqref{eq:metric} near the poles
by setting $\theta = \theta_{\pm} \pm \rho$
(with $\theta_+=0$ and $\theta_-=\pi$) near each axis:
\be
ds^2 \sim \frac{1}{H^2}\frac{\Sigma r^2}{g(\theta_{\pm})}
\bigg[d\rho^2 +  \frac{g^2(\theta_{\pm}) \rho^{2}}{K^2}d\phi^2\bigg]\,.
\ee
The deficit on each axis is then read off as:
\be\label{eq:deficits}
\delta_\pm=2\pi\bigg[1-\frac{g(\theta_\pm)}{K}\bigg]
= 2\pi\bigg[1-\frac{\Xi \pm 2mA}{K}\bigg]\,.
\ee
If $A=0$, both deficits are identical and can be interpreted as a cosmic
string through the black hole \cite{Aryal:1986sz,Achucarro:1995nu} of tension
\be
\mu = \frac{\delta}{8\pi} = \frac14 \left [1-\frac{\Xi}{K}\right]\,.
\label{mudef}
\ee
If $A$ is nonzero however, then there is an asymmetry in the spacetime,
with differing deficits at north and south poles:
\be
\mu_\pm = \frac14 \left [1-\frac{\Xi \pm 2\tilde{m}} {K}\right]\,,
\ee
that produces a nett force on the black hole, hence acceleration.

It is now evident that if we choose $K$ to obtain a particular value of the
conical deficit on one axis, that choice of $K$ has a global impact:
$A$ then regulates the distribution of tensions between the axes.
It is also worth mentioning that although a negative deficit (otherwise
known as an excess) is possible, it would be sourced by a negative
energy object and hence in general associated with instabilities
(though see \cite{Costa:2000kf, Herdeiro:2009vd, Herdeiro:2010aq, Krtous:2019fpo}).
We therefore restrict ourselves
to positive energy sources, thus (taking $A>0$ without loss of generality)
$K\geqslant \Xi+2\tilde{m}$. In most of the literature
on accelerating black holes, the deficit along one axis (here, the north)
is chosen to vanish, i.e.\ $K= \Xi + 2\tilde m$. However, we will not
make this restriction here, unless stated explicitly.

\subsection{Thermodynamics of accelerated black holes}

The properties of slowly accelerating black holes have been studied in
recent years and our understanding of their thermodynamics has greatly
improved over time \cite{Appels:2016uha,Appels:2017xoe,Gregory:2017ogk,
Anabalon:2018ydc}. The full thermodynamics for the general accelerating
black hole is given by the extended first law \cite{Appels:2017xoe}:
\be
\delta M  = T \delta S + \Phi \delta Q + \Omega \delta J + V \delta P
+ \lambda_{+}\delta \mu_{+} + \lambda_{-}\delta \mu _{-}\,,
\label{eq:First_law}
\end{equation}
where the enthalpy is
\be
M = \frac{ m \ep{\Xi + a^2 /  \ell^2} \ep{ 1-A^2 \ell^2 \Xi} }
{K\Xi\alpha \ep{1+a^2 A^2} }\,,
\label{originalM}
\ee
(with $\alpha$ defined in \eqref{alphadef}) and the
six thermodynamic charges $S,Q,J,P,\mu_{\pm}$ together with
their corresponding potentials $T,\Phi,\Omega,V,\lambda_{\pm}$
are given in terms of the six black hole parameters $A,a,m,e,\ell,K$
as  \cite{Anabalon:2018qfv}
\be
\beal
T&= \frac{f'_+ r_+^2}{4\pi\alpha(r_+^2+a^2)}\,, \quad
S=\frac{\pi(r_+^2+a^2)}{K(1-A^2r_+^2)}\,,\\
J& =\frac{ma}{K^2}\,,  \quad \Omega=  \Omega_H-\Omega_\infty\,
=\left ( \frac{Ka}{\alpha(r_+^2+a^2)}\right ) -
\left ( -\frac{aK(1-A^2\ell^2\Xi)}{\ell^2\Xi \alpha(1+a^2A^2)}\right) \,,\\
Q&= \frac{e}{K}\,,\quad \Phi=\Phi_t=\frac{er_+}{(r_+^2+a^2)\alpha}\,,\\
P &= \frac{3}{8\pi \ell^2} \,,\;\;
V = \frac{4\pi}{3K\alpha} \left [ \frac{r_+(r_+^2 + a^2)}{(1-A^2 r_+^2)^{2}}
+ \frac{m[a^2 + A^2 \ell^4 \Xi^2]}
{(1+a^2 A^2) \Xi} \right]\,,\\
\lambda_\pm &= \frac{-r_+}{\alpha(1\pm Ar_+)} +\frac{m}{\alpha}
\frac{[\Xi +   \frac{a^2}{\ell^2} (2-A^2\ell^2 \Xi)]}{(1+a^2 A^2)\Xi^2}
\pm \frac{A \ell^2 (\Xi +  a^2/\ell^2 )}{\alpha(1+a^2A^2)}\,.
\eeal
\label{TDcharges}
\ee
These charges also satisfy a Smarr relation \cite{Smarr:1972kt}
\begin{equation}
\label{eq:Smarr}
M= 2(TS + \Omega J - PV ) + \Phi Q\,.
\end{equation}
A description of how the potentials were obtained, using both conformal and
holographic techniques is given in Anabalon et al.\ \cite{Anabalon:2018qfv}.

Despite the fact that the tensions $\mu_{\pm}$ are natural variables, and indeed
correspond to physical objects (cosmic strings emerging from the event
horizon \cite{Achucarro:1995nu,Gregory:1995hd}), expressing the
charges and potentials in terms of extensive variables
\cite{Gregory:2019dtq} reveals that the thermodynamics is more naturally
delineated into an overall and differential
conical deficit, $\Delta$ and $C$ respectively:
\be
\beal
\Delta &= 1 - 2(\mu_{+} + \mu_{-}) = {\Xi \over K}\,, \\
C &= {\mu_{-} - \mu_{+} \over \Delta} = {\tilde m \over \Delta K} = { m A \over \Xi}\,.
\eeal
\label{delta-c}
\ee
Since the tensions are bounded from below by the positivity of energy,
and above by the maximum conical deficit of $2\pi$, we have
\be
0 \leqslant \mu_{+} \leqslant \mu_{-} \leqslant {1/4}\,,
\ee
which translates into bounds for $C$:
\be
0 \leqslant C \leqslant \min\left\{{1 \over 2},{1-\Delta \over 2\Delta}  \right\}\,.
\ee

The Christodulou-like formula for the enthalpy then reads \cite{Gregory:2019dtq}
\be\label{Christodulou}
M^2 =
\frac{\Delta S}{4\pi}
\Bigl[ \left(1+\frac{\pi Q^2}{\Delta S}+\frac{8PS}{3\Delta}\right)^2
+\left(1+\frac{8PS}{3\Delta}\right)
\left(\frac{4\pi^2 J^2}{\Delta^2S^2} - \frac{3C^2\Delta}{2PS}\right)
\Bigr]\,,
\ee
while the other expressions are
\be
\beal
V &=
\frac{2S^2}{3\pi M} \left[ 1+\frac{\pi Q^2}{\Delta S}+\frac{8PS}{3\Delta}
+ \frac{2\pi^2J^2}{(\Delta S)^2} +\frac{9C^2\Delta^2}{32P^2S^2}
\right] \,,\\
T &= \frac{\Delta}{8\pi M}
\Bigg[\left(1+\frac{\pi Q^2}{\Delta S}+\frac{8PS}{3\Delta}\right)
\left(1-\frac{\pi Q^2}{\Delta S}+\frac{8PS}{\Delta}\right )
-\frac{4\pi^2 J^2}{(\Delta S)^2} -4C^2
\Bigg]
\,,\\
\Omega &= \frac{\pi J}{S M \Delta} \left(1+\frac{8PS}{3\Delta}\right) \,,\\
\Phi &= \frac{Q}{2M}\left(1+\frac{\pi Q^2}{S\Delta}+\frac{8PS}{3\Delta}\right)\,,\\
\lambda_\pm    &=
\frac{S}{4\pi M}
\Bigg[ \! \!\left ( \!\frac{8PS}{3\Delta} + \frac{\pi Q^2}{\Delta S} \!\right)^2
\!\!+ \frac{4\pi^2 J^2}{(\Delta S)^2} \left (\!1+ \frac{16PS}{3\Delta}\!\right)
\!-\left ( 1 \mp 2C \right)^2 \pm \frac{3C\Delta}{2PS}\!
\Bigg]
\,,
\eeal
\label{TDSPJ}
\ee
or considering the conjugates to $\Delta$ and $C$ instead,
\be\label{alt}
\beal
\lambda_\Delta &= - \frac{S}{8\pi M} \Bigg[ \!
\!\left ( \!\frac{8PS}{3\Delta} + \frac{\pi Q^2}{\Delta S} \!\right)^2
\!\!+ \frac{4\pi^2 J^2}{(\Delta S)^2} \left (\!1+ \frac{16PS}{3\Delta}\!\right)
- 1 + C^2 \left ( 4 - \frac{3\Delta}{PS}\! \right)
\Bigg]\,,\\
\lambda_C &= - \frac{\Delta C S}{\pi M} \left [ 1 + \frac{3\Delta}{4PS} \right]\,.
\eeal
\ee

These expressions, \eqref{Christodulou}, \eqref{TDSPJ}, and \eqref{alt}
are most useful for exploring the
general thermodynamical properties of the black holes, however, we will refer
to the parametric expressions \eqref{TDcharges} when discussing the
ultraspinning black hole.

\subsection{Reverse Isoperimetric Inequality}

The fact that the thermodynamic quantities of the accelerated black
holes obey the {\em Reverse Isoperimetric Inequality} \cite{Cvetic:2010jb}
(roughly, a statement that black holes like to be round) has been shown in
\cite{Gregory:2019dtq}. Let us repeat here the corresponding argument.

Squaring the formula \eqref{TDSPJ} for $V$, we have
\be
\Bigl(\frac{3V}{4\pi}\Bigr)^2\Bigl(\frac{\pi}{S}\Bigr)^3
=\frac{S}{4\pi}\frac{1}{M^2}\left[ 1+\frac{\pi Q^2}{\Delta S}+\frac{8PS}{3\Delta}
+ \frac{2\pi^2J^2}{(\Delta S)^2} +\frac{9C^2\Delta^2}{32P^2S^2}
\right]^2\,.
\ee
Thence, upon using the Christodulou formula \eqref{Christodulou}
to eliminate $M$, this yields
\bea
\Delta\Bigl(\frac{3V}{4\pi}\Bigr)^2\Bigl(\frac{\pi}{S}\Bigr)^3&=&
\frac{\left[ 1+\frac{\pi Q^2}{\Delta S}+\frac{8PS}{3\Delta}
+ \frac{2\pi^2J^2}{(\Delta S)^2} +\frac{9C^2\Delta^2}{32P^2S^2}\right]^2}
{\Bigl[ \left(1+\frac{\pi Q^2}{\Delta S}+\frac{8PS}{3\Delta}\right)^2
+\left(1+\frac{8PS}{3\Delta}\right)
\left(\frac{4\pi^2 J^2}{\Delta^2S^2} - \frac{3C^2\Delta}{2PS}\right)
\Bigr]}\nonumber\\
&\geq&
\frac{\left[ 1+\frac{\pi Q^2}{\Delta S}+\frac{8PS}{3\Delta}
+ \frac{2\pi^2J^2}{(\Delta S)^2}\right]^2}{\Bigl[\left(1+\frac{\pi Q^2}{\Delta S}
+\frac{8PS}{3\Delta}\right)^2
+2\left(1+\frac{\pi Q^2}{\Delta S}+\frac{8PS}{3\Delta}\right)
\frac{2\pi^2 J^2}{\Delta^2S^2}\Bigr]}
\geq 1\,.
\eea
We have thus verified the refined Reverse Isoperimetric Inequality
\cite{Gregory:2019dtq}
\be\label{ISO}
\Bigl(\frac{V}{V_0}\Bigr)^2\geq \frac{1}{\Delta}\Bigl(\frac{A}{A_0}\Bigr)^3\,,
\ee
where $V_0$ and $A_0$ are the volume and area of a unit ball,
$V_0=\frac{4}{3}\pi$ and $A_0=4\pi$,
and the inequality is saturated if and only if $C=0=J$.

\section{Critical black holes and their thermodynamics}
\label{sec:critTD}

\subsection{Critical limit}

Having discussed the slowly accelerating C-metric, and the parametric
restrictions that this geometry requires, we now turn to the critical black
holes we are interested in exploring.

The term \emph{critical} is used to describe a geometry in which at
least one of the tensions has its maximal value of $1/4$, i.e., where the
deficit becomes $2\pi$ as in the ultra-spinning black hole.
For the ultraspinning Kerr-AdS black hole, this corresponds to
saturating an upper bound on rotation, however, in our accelerating
black hole metric, the deficit along one axis can become $2\pi$, even
in the absence of rotation, e.g.\ for $mA=1/2$ in the `black bottles'
of \cite{Chen:2016rjt, Chen:2016jxv}. We can therefore think of criticality
as saturation of an upper bound for the mass parameter $\tilde{m}$,
\be
mA = \Xi/2\,.
\label{criticalm}
\ee
\begin{figure}
\centering
\begin{subfigure}[b]{0.25\textwidth}
\includegraphics[width=\textwidth]{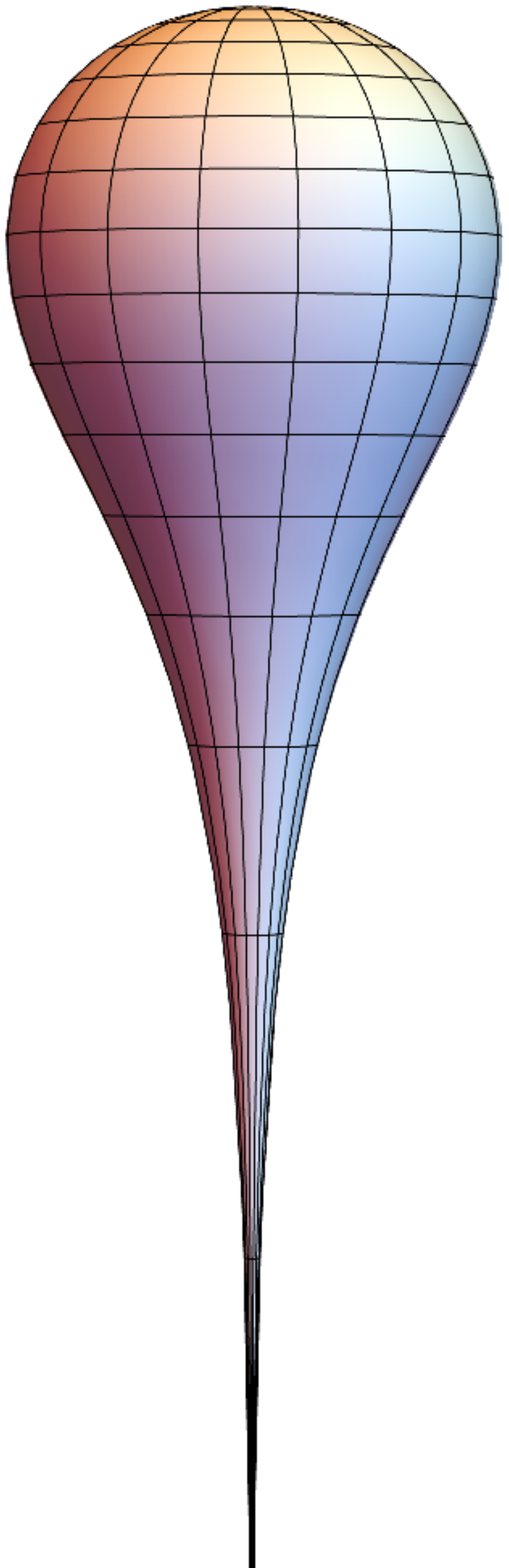}
\caption{\label{fig:embed3Dus-smooth}}
\end{subfigure}\qquad
\begin{subfigure}[b]{0.25\textwidth}
\includegraphics[width=\textwidth]{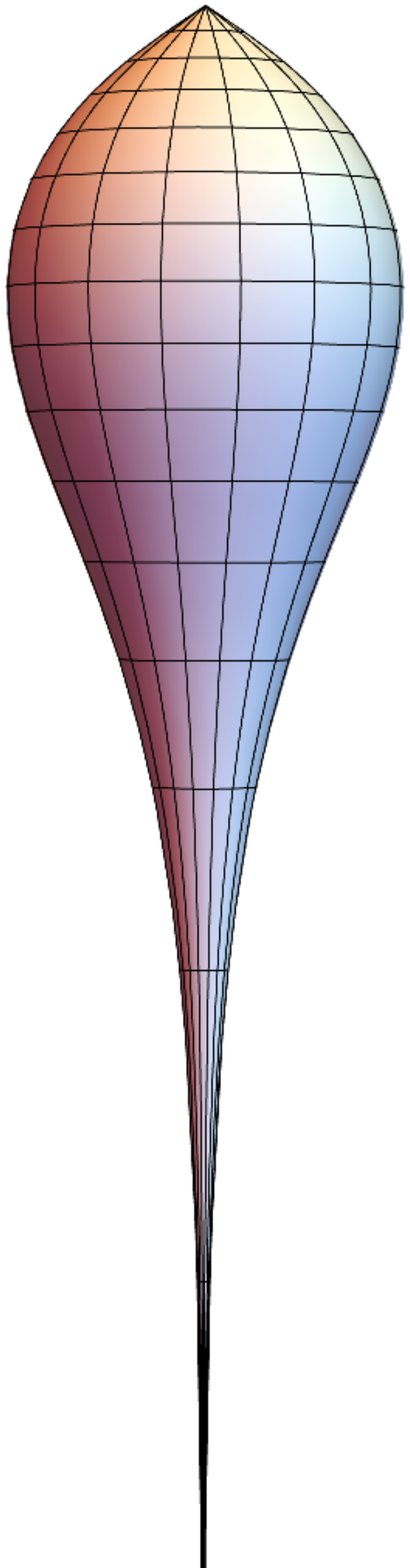}
\caption{\label{fig:embed3Dus-Kbig}}
\end{subfigure}\qquad
\begin{subfigure}[b]{0.25\textwidth}
\includegraphics[width=\textwidth]{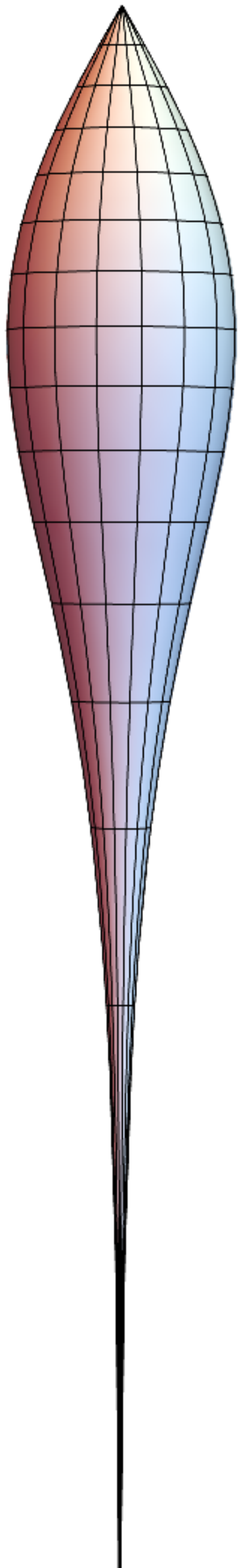}
\caption{\label{fig:embed3Dus-Kbigger}}
\end{subfigure}
\caption{
Horizon embedding of critical black holes. We display the
horizon embeddings in $\mathbb{R}^3$ of the critical C-metrics for
$m=9\ell$, $A=0.04\ell$ and (a) $K=2\Xi$ (b) $K=2.4\Xi$ (c) $K=4\Xi$.}
\label{fig:embed3Dus}
\end{figure}
With this choice, the south pole axis is effectively removed from the
spacetime, while the north pole axis may still have a conical deficit,
determined by the ratio of $K/\Xi$, see Fig.~\ref{fig:embed3Dus},
where the embedding of the event horizon of a critical black hole is
displayed for various such ratios.

\begin{figure}
\centering
\includegraphics[width=\textwidth]{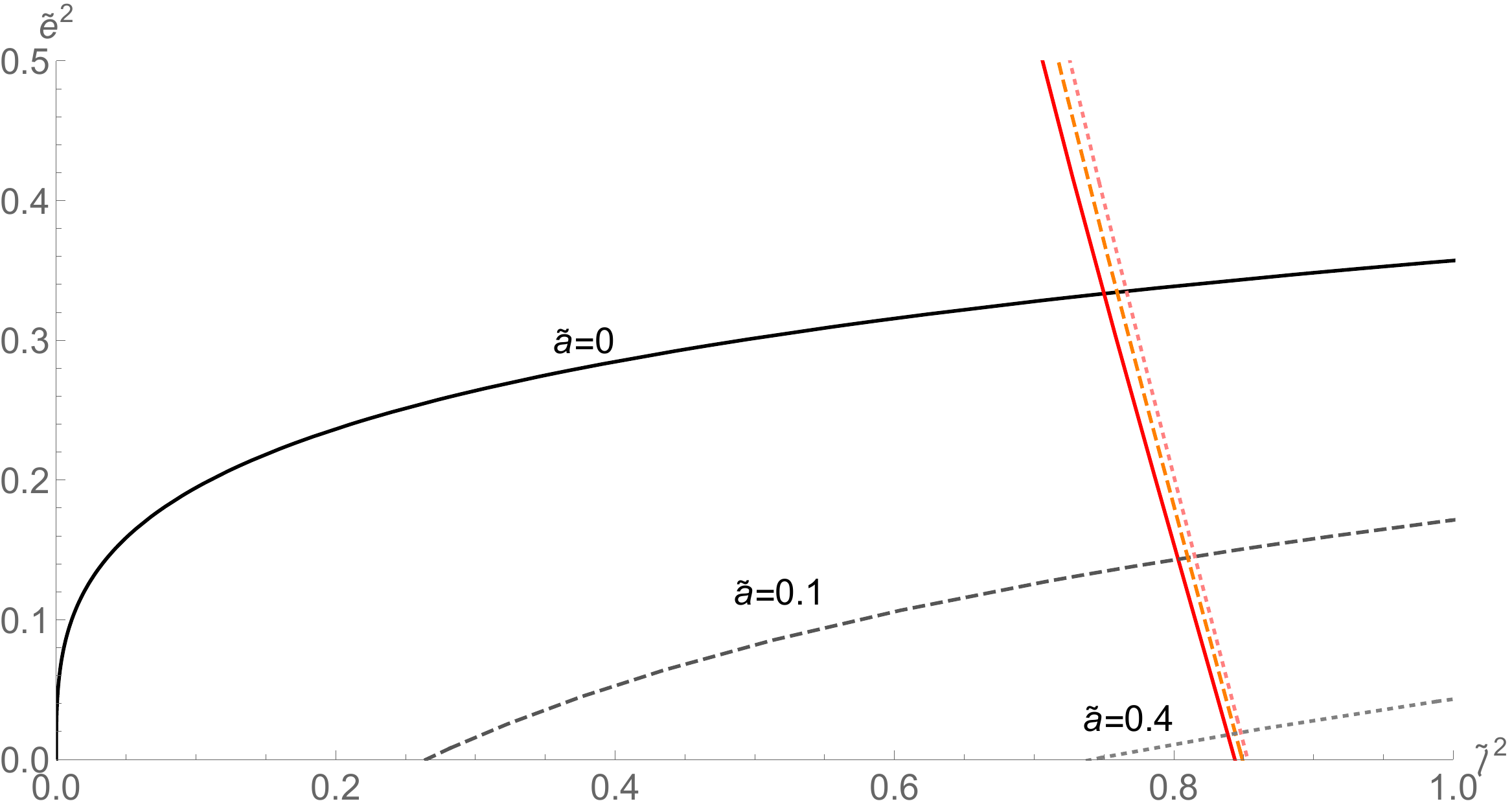}
\caption{The allowed values of $\tilde \ell$ and $\tilde e$:
The upper bound for $\tilde{e}$ from extremality is shown in black/grey, and
the upper bound for $\tilde{\ell}$ from the slow acceleration limit is shown in
red/pink for sample values of $\tilde a$ as labelled. The upper bound for
$\tilde a$ is $\tilde{a}^2 = 3-2\sqrt{2}$.}
\label{fig:aA}
\end{figure}

Since in the process of taking the critical limit, only one parameter is
eliminated,  by imposing \eqref{criticalm}, we have a
three-parameter family of critical accelerating black
holes, parametrised by $\tilde{e} = eA$, $\tilde{a} = aA$, and $\tilde{\ell} = A\ell$,
with the mass given by \eqref{criticalm}. Once again, these parameters
are constrained by $g(\theta)$, and the
slow-acceleration/extremal limits for the black hole:
\be
\beal
\text{extremal limit} &
\begin{cases}
\beal
{\tilde\ell}^2_{\text{ext}} &= \frac{{\tilde a}^2 + 3 {\tilde a}^2 {\tilde r}_+
+ 4 {\tilde r}_+^3 + {\tilde r}_+^4-{\tilde r}_+^5}{(1-{\tilde r}_+)^3(1+{\tilde r}_+)^2}\,,\\
{\tilde e}^2_{\text{ext}} &= \frac{-{\tilde a}^4 - 3 {\tilde a}^4 {\tilde r}_+
+ 2 {\tilde a}^2 {\tilde r}_+^2 - 2 {\tilde a}^2 {\tilde r}_+^3 +3 {\tilde r}_+^4+{\tilde r}_+^5}
{{\tilde a}^2 + 3 {\tilde a}^2 {\tilde r}_+ + 4 {\tilde r}_+^3 + {\tilde r}_+^4 - {\tilde r}_+^5}\,,
\eeal
\end{cases}\\
\text{slow acc.\ limit}&
\begin{cases}
\beal
{\tilde\ell}^2_{\text{acc}} &=
\frac{1 + y_+ - 4 y_+^2 - 3 {\tilde a}^2 y_+^4
+ {\tilde a}^2 y_+^5} {(1-y_+)^2(1+y_+)^3}\,,\\
{\tilde e}^2_{\text{acc}} &= \frac{-1 + 3y_+ +2 {\tilde a}^2 y_+^2
+ 2 {\tilde a}^2 y_+^3+3 {\tilde a}^4 y_+^4 - {\tilde a}^4 y_+^5}
{1+y_+-4y_+^2-3{\tilde a}^2 y_+^4+{\tilde a}^2 y_+^5}\,.
\eeal
\end{cases}
\eeal
\ee
See Fig.~\ref{fig:aA} for a plot of parameter space.
Note, the constraint from $g(\theta)$ is automatically (marginally) satisfied due
to the choice of $\tilde{m}$.

\subsection{Thermodynamics and absence of superentropicity}

The above constructed critical black holes $(\mu_-=1/4)$
were simply obtained by setting $2 mA = \Xi$, that is,
\be\label{crit}
C=\frac12\,,\quad \Delta=\frac{1}{2}-2\mu_+\,.
\ee
The associated allowed range for $\Delta$ is $\Delta \in [0,1/2]$,
with the lower (upper) bound corresponding to the
upper (lower) value that $\mu_{+}$ can take; thus
$\mu_+ = 0 \longleftrightarrow \Delta = 1/2$, and $\mu_+ \to 1/4 \longleftrightarrow
\Delta \to0$.

The criticality condition \eqref{crit} is hard to impose at the
parametric level, \eqref{TDcharges}, but very simple for the expressions
\eqref{Christodulou}, \eqref{TDSPJ}, and \eqref{alt}.
The limit is smooth and simply yields
\be
\beal
\label{MVT_Penrose}
M^2 &=
\frac{\Delta S}{4\pi}
\Bigl[ \left(1+\frac{\pi Q^2}{\Delta S}+\frac{8PS}{3\Delta}\right)^2
+\left(1+\frac{8PS}{3\Delta}\right)
\left(\frac{4\pi^2 J^2}{\Delta^2S^2} - \frac{3\Delta}{8PS}\right)
\Bigr]\,,\\
V &=
\frac{2S^2}{3\pi M} \left[ 1+\frac{\pi Q^2}{\Delta S}+\frac{8PS}{3\Delta}
+ \frac{2\pi^2J^2}{(\Delta S)^2} +\frac{9\Delta^2}{128P^2S^2}
\right] \,,\\
T &= \frac{\Delta}{8\pi M}
\Bigg[\left(1+\frac{\pi Q^2}{\Delta S}+\frac{8PS}{3\Delta}\right)
\left(1-\frac{\pi Q^2}{\Delta S}+\frac{8PS}{\Delta}\right )
-\frac{4\pi^2 J^2}{(\Delta S)^2} -1
\Bigg]
\,,\\
\Omega &= \frac{\pi J}{S M \Delta} \left(1+\frac{8PS}{3\Delta}\right) \,,\\
\Phi &= \frac{Q}{2M}\left(1+\frac{\pi Q^2}{S\Delta}+\frac{8PS}{3\Delta}\right)\,,\\
\lambda_\Delta &= - \frac{S}{8\pi M} \Bigg[ \!
\!\left ( \!\frac{8PS}{3\Delta} + \frac{\pi Q^2}{\Delta S} \!\right)^2
\!\!+ \frac{4\pi^2 J^2}{(\Delta S)^2} \left (\!1+ \frac{16PS}{3\Delta}\!\right)
 - \frac{3\Delta}{4PS}
\Bigg]\,.
\eeal
\ee
These quantities obey the full cohomogeneity first law,
\be
\delta M  = T \delta S + \Phi \delta Q + \Omega \delta J
+ V \delta P+\lambda_\Delta \delta \Delta\,,
\label{eq:First_lawCrit}
\end{equation}
together with the corresponding Smarr relation \eqref{eq:Smarr}.

Of course, the proof of the reverse isoperimetric inequality \eqref{ISO}
presented above for accelerated black holes goes through exactly the
same way for their critical subfamily and this is despite the fact that the
horizon of critical black holes is non-compact (as is the horizon of
ultraspinning black holes). Note also that since $C=1/2$ the inequality
can no longer be saturated and is a strict inequality.

Also, note that by taking another limit, $\Delta\to 0$, one
formally obtains a critical black hole with maximal conical
deficits on both poles, that in fact is the superspinning black hole.
However, it is obvious from the expressions
\eqref{MVT_Penrose} above, that the thermodynamic quantities
such as mass $M$, and angular velocity $\Omega$ diverge in this
limit. Therefore, either one accepts that the thermodynamics is ill-defined
in this limit,
or one looks for new (renormalised) thermodynamic parameters.

\section{Comparison to ultraspinning black holes} \label{sec:comp}

\subsection{The superentropic argument}

Let us now compare the critical limit to the ultraspinning limit of the
Kerr-AdS spacetime.
The charged Kerr-AdS black holes are obtained by setting $A=0$
in the metric \eqref{eq:metric}. For simplicity, and without loss
of generality for the purposes of this discussion, we
will also take the uncharged limit $e=0$, so that the metric becomes
\be
ds^2 = \frac{f(r)}{\Sigma}\Big[
{dt}-a\,\sin^2\theta \frac{ d\phi}{K} \Big]^2
- \frac{\Sigma dr^2}{f(r)} \; - \frac{\Sigma r^2 d\theta^2}{g(\theta)}
- \frac{g(\theta)\sin^2\theta}{\Sigma r^2} \Big[{adt}-(r^2+a^2)
\frac{d\phi}{K}\Big]^2\,,
\label{supermetric}
\ee
where we now have
\be
\beal
f(r) &=\left[1-\frac{2m}{r}+\frac{a^2}{r^2}\right]
+\frac{r^2+a^2}{\ell^2}\,,\quad
g(\theta) =1+ (\Xi-1)\cos^2\theta\,,\\
\Sigma &=1+\frac{a^2}{r^2}\cos^2\theta\,, \qquad
\Xi = 1 -\frac{a^2}{\ell^2}\,.
\eeal
\ee
In the previous literature, one sets $K\equiv\Xi$, so that
there is no conical deficit in the spacetime. The thermodynamics of these
black holes was worked out definitively in \cite{Caldarelli:1999xj,Gibbons:2004ai},
the key insight being that the boundary has a non-zero angular velocity,
\be\label{Ominf1}
\Omega_\infty = \lim_{r\to\infty} -\frac{g_{t\phi}}{g_{\phi\phi}}
= - \frac{aK}{\ell^2\Xi}\,,
\ee
implying that the total angular velocity ought to be re-normalised,
$\Omega=\Omega_H-\Omega_\infty$. Further, a computation of the
mass of the spacetime, using an appropriately normalised Killing
vector, $\partial_t-\Omega_\infty \partial_\phi$, yielded $M=m/\Xi^2$
for the enthalpy. These results are entirely consistent with
\eqref{originalM}, \eqref{TDcharges}, once one sets $K=\Xi$.
Crucially, when considering a varying $\Lambda$, the inclusion of these
normalisations for enthalpy and angular velocity leads to
an enthalpy dependent correction term in the thermodynamic volume:
\be
V = V_0+V_1 = \frac{4\pi r_+(r_+^2 + a^2)}{3K} + \frac{4\pi M a^2}{3}\,.
\label{v0v1}
\ee

The ultra-spinning limit is obtained by taking the limit in which
$a\to\ell$ ($\Xi\to 0)$, but because of the identification of $K$ with $\Xi$, this
results in an apparently singular metric. This was resolved in
\cite{Klemm:2014rda,Hennigar:2014cfa} by defining a new angular
coordinate, $\psi = \phi/\Xi$, so that $\psi$ formally becomes noncompact in the
ultra-spinning limit. This new angular coordinate is then given a finite
range, $\Delta \psi = \mu_S$. Since $g(\theta) \to \sin^2\!\theta$ the
limit yields the {\em Superentropic} black hole
\bea
ds^2 &=& \frac{f(r)}{\Sigma}\Big[
{dt}-\ell\,\sin^2\theta d\psi \Big]^2
- \frac{\Sigma dr^2}{f(r)} \; - \frac{\Sigma r^2 d\theta^2}{\sin^2\!\theta}
- \frac{\sin^4\!\theta}{\Sigma r^2} \Big[{\ell dt}-(r^2+\ell^2)
d\psi\Big]^2\,,\nonumber
\label{supermetric}\\
f(r) &=&\frac{\ell^2}{r^2}\left(1+\frac{r^2}{\ell^2}\right)^2-\frac{2m}{r}\,,\quad
\Sigma =1+\frac{\ell^2}{r^2}\cos^2\!\theta\,,
\eea
which was assigned the following thermodynamic parameters
\cite{Klemm:2014rda,Hennigar:2014cfa}:
\bea
M_S &=& \frac{\mu_S m}{2\pi}\;, \quad
S_S = \frac{\mu_S}{2} (r_+^2+\ell^2)\;,\quad
T_S = \frac{f'(r_+) r_+^2}{4\pi(r_+^2+\ell^2)}\,,\quad J_S = M_s\ell\;, \nonumber\\
\Omega_S &=& \frac{\ell}{r_+^2+\ell^2}\;, \quad
V_S = \frac{2\mu_S r_+}{3} (r_+^2+\ell^2)\,,\quad
\lambda_S =  \frac{m}{4\pi} \frac{(\ell^2-r_+^2)}{(r_+^2+\ell^2)}\,,
\label{TDsuper}
\eea
where the subscript $S$ is used to denote these specific `superentropic'
definitions, and we have relabelled the thermodynamic length parameter,
denoted $K$ in \cite{Hennigar:2014cfa} as $\lambda_S$, dual
to a variation of the parameter $\mu_S$,
\be\label{supFirst}
\delta M_S=T_S \delta S_S+\Omega_S \delta J_S+V_S \delta P
+\lambda_S \delta \mu_S\,.
\ee
The first law is obviously degenerate as only 3 parameters,
$\{r_+, \ell, \mu_S\}$, can be varied independently;
the mass $M_S$ and angular momentum $J_S$ charges obey
the `chirality condition' ${J_S}={M_S}\ell$.

Note that $\Omega_S$ is simply the angular velocity, defined
by $\omega=-g_{t\psi}/g_{\psi\psi},$ evaluated on the horizon,
$\Omega_S=\omega(r_+)$, while the corresponding quantity at
infinity diverges. That is, in  \cite{Hennigar:2014cfa} the authors
set formally $\Omega_\infty=0$, there is no renormalisation of
angular velocity, nor of the timelike Killing vector, and in
consequence, there is no adjustment
of the enthalpy `$M$', nor a  correction to the thermodynamic
volume. As a result, the volume is simply the geometric volume, thus
the standard mathematical {\it Isoperimetric} inequality applies, and the
entropy is now minimised by the contained volume. This fascinating
result has caused some puzzlement, as the thermodynamic parameters
\eqref{TDsuper} are not obtained as an ``$a\to\ell$'' limit of the
conventional parameters, \eqref{TDcharges},
nor does it seem possible to obtain one of these black holes by some sort
of continuous process. In addition, the idea that the entropy can be
unbounded for a fixed volume suggests that superentropic black
holes should be somehow unstable, a notion explored (in a different
context) by Johnson \cite{Johnson:2019mdp}, see also
\cite{Cong:2019bud,Johnson:2019wcq}. Thus the {\em Superentropic}
black hole is worthy of further study.

One of the problems of the thermodynamic parameters of \cite{Hennigar:2014cfa}
is that setting $a\equiv \ell$ means that the angular momentum and
thermodynamic pressure are no longer independent variables.
In other words, the first law no longer has full cohomogeneity.
Further, the discrete alteration of the periodicity of the angular coordinate
is equivalent to a sudden shift of the conical deficit from $0$ to $2\pi$, as
one is setting $K=\Xi$ for the sub-rotating black holes (giving
$\mu=0$) but for $a=\ell$, the periodicity of the original $\phi$ coordinate,
set to $\mu\Xi$ by Hennigar et al.\ \cite{Hennigar:2014cfa}, now vanishes.
However, since we have a set of thermodynamic variables that include
potential variations in the conical deficit, we can now examine this
superentropic ultra-spinning limit afresh, and try to understand what
lies behind this phenomenon.

\subsection{Kerr-AdS with conical deficits}

Let us return to the general metric \eqref{supermetric}, retaining the
parameter $K$, and re-examine the thermodynamics of the
ultraspinning Kerr in the light of allowing for conical deficits.
First, note that in the limit $a\to\ell$,
$g(\theta) \to \sin^2\theta$, thus defining a new angular coordinate
\be
\Theta = \log \tan \frac{\theta}{2}\,,
\ee
the angular part of our metric in terms of $\Theta$ and $\phi$ is
manifestly non-compact. The parameter $K$ however now has no
apparent physical meaning, as the deficit along the axis, defined
by the tension \eqref{mudef} becomes maximal:
\be
\delta =  8\pi \mu = 2\pi \left [ 1 - \frac{\Xi}{K} \right ] \to 2\pi\,.
\ee
However, guided by the discussion in \cite{Hennigar:2014cfa},
define
\be
\mu_S = \frac{2\pi}{K} = \frac{2\pi}{\Xi} (1-4\mu)\,,
\label{musdef}
\ee
that will play the role of a ``spectator tension''.

We now re-derive the thermodynamics of the superspinning black hole
by taking a continuous limit of the generic, fully cohomogeneous, variables
given in \eqref{TDcharges} by approaching the limit $a\to\ell$ from a
more continuous perspective, taking a family of black holes with
$a/\ell = \sqrt{1-\Xi}$ fixed, then taking the limit $\Xi\to0$.

Imposing the constraint that $\Xi$ is a constant means that
$J$ and $P$ are no longer independent thermodynamic variables
($\delta a = a \delta \ell/\ell$),
thus keeping a first law with variations of both angular momentum
and pressure is a bit disingenuous, and such a first law no longer
has full cohomogeneity. Instead, the variation of the angular momentum
yields contributions to the pressure variation, as well as terms that
contribute to the enthalpy.
\be
\delta J = a\delta \left (\frac{m}{K^2} \right)+ \frac{m}{K^2} \delta a
= a \delta \left (\frac{m}{K^2} \right) - \frac{4\pi\ell^2 }{3} \frac{ma}{K^2} \delta P\,.
\label{varyJ}
\ee
Therefore, in deriving the superentropic variables \eqref{TDsuper},
Hennigar et al.\ have effectively used this equivalence between
the variation of $J$ and $P$ to ``re-organise'' terms in the
thermodynamic potentials,
and (roughly) the angular momentum subtraction at infinity term
cancels off the compensating thermodynamic volume term to
yield an uncorrected $V$, hence a standard Isoperimetric inequality.

To track through the play-off between the various terms, start with the
first law
\be
\delta M = T\, \delta S + \Omega\, \delta J + V\, \delta P
+ 2 \lambda\, \delta \mu\,,
\ee
with the thermodynamic variables pertinent to the discussion being:
\be
\beal
M&= \frac{m}{K\Xi}\,, \quad
J =\frac{ma}{K^2}\,,  \quad \Omega=  \Omega_H-\Omega_\infty\,
= \frac{Ka}{(r_+^2+a^2)} + \frac{aK}{\ell^2\Xi} \,,\\
V &= V_0 + V_1 = \frac{4\pi r_+(r_+^2 + a^2)}{3K}+\frac{4\pi m a^2}{3K\Xi}\,,
\quad \lambda = -r_+ +\frac{m}{\Xi^2}
\left ( 1 + \frac{a^2}{\ell^2}\right) \,.
\eeal
\ee

Now, using the fact that $\Xi$ is a constant, and noting the definition of
$\mu_S$ above, \eqref{musdef}, we see that
\be
\delta M = \frac1\Xi \,\delta (\frac{m}{K}) =  \frac1\Xi \,\delta (\frac{\mu_s m}{2\pi})
= \frac1\Xi \,\delta M_S\,,
\label{MMS}
\ee
where $M_S$ is the superentropic variable in \eqref{TDsuper}, but
now defined at {\it finite} $\Xi$. We can also relate the variation of tension
to the rescaled `spectator' tension $\mu_S$:
\be
\delta \mu = - \frac{\Xi}{8\pi} \delta \mu_S\,.
\ee

Finally, looking at the angular momentum, and using
\eqref{musdef}, \eqref{varyJ}, \eqref{MMS} we see that
$J = aM_s/K$, and we can write the variation of $J$ in two
useful alternate forms:
\be
\beal
\delta J &= \frac{a}{K} \delta M_S
+ \frac{am}{K} \frac{\delta \mu_S}{2\pi}  - \frac{4\pi\ell^2 }{3} \frac{ma}{K^2} \delta P\\
&= \frac{1}{K} \delta (a M_S )
+ aM_S \frac{\delta \mu_S}{2\pi}\,,
\eeal
\label{Jvary}
\ee
thus using the first expression in $\Omega_\infty \delta J$, and the
second in $\Omega_H \delta J$, we see that
\be
\Omega \delta J = \frac{a^2}{\ell^2\Xi} \delta M_S +\frac{a}{r_+^2+a^2} \delta (a M_S )
+ \left [ \frac{a m }{r_+^2+a^2}
+ \frac{a^2m}{\ell^2\Xi} \right] \frac{\delta \mu_S}{2\pi}
- \frac{4\pi}{3} \frac{ma^2}{K\Xi} \delta P\,.
\ee
Note, the last piece in the above expression is in fact $-V_1\delta P$, so we
now see how the compensating term in the thermodynamic volume
that maintains the reverse isoperimetric inequality is cancelled.

Putting together all these pieces, we see our first law becomes
\be
\beal
\delta M_S &= T\, \delta S +\frac{a}{r_+^2+a^2} \delta (a M_S )
+ V_0\, \delta P
+ \left [ \frac{a m }{r_+^2+a^2}
+ \frac{a^2m}{\ell^2\Xi} - \frac{\lambda\Xi}{2} \right] \frac{\delta \mu_S}{2\pi}\\
&= T\, \delta S +\frac{a}{r_+^2+a^2} \delta (a M_S )
+\frac{2\mu_S r_+}{3} (r_+^2+a^2) \, \delta P
+\left [ r_+\Xi - m \frac{r_+^2-a^2}{r_+^2+a^2} \right] \frac{\delta \mu_S}{4\pi}\,.
\eeal
\ee
We see a clear parallel with the thermodynamic variables of \eqref{TDsuper},
indeed, defining a new angular momentum charge, potential, and
thermodynamic length
\be
J_S = aM_S\,,\quad
\Omega_S = \frac{a}{r_+^2+a^2}\,,\quad
\lambda_S = \frac{1}{4\pi}\Bigl( r_+\Xi - m \frac{r_+^2-a^2}{r_+^2+a^2}\Bigr)\,,
\ee
which are identical to those in \eqref{TDsuper} when $a=\ell$, we do indeed
confirm the consistency of the Hennigar et al.\ first law, \eqref{supFirst}, but
now for {\it finite} $\Xi$. We stress that although consistent for
any finite $\Xi$, such a first law is not the correct one -- it uses the
wrong thermodynamic quantities, e.g., $\Omega_S$ lacks the
contribution from rotation at infinity.
Note also that after the limit $a\to \ell$, the spectator tension
should really not be varied (the parameter $K$ is no longer physical)
and the term $\lambda_S \delta \mu_S$ in \eqref{supFirst} should be omitted.

However, studying the system at finite $\Xi$ reveals something
interesting that was missed in \cite{Hennigar:2014cfa}.
Conventionally, consistency of thermodynamics has been used as the principle
upon which to define the thermodynamic parameters,
and provided there is full cohomogeneity this seems to be correct. However, once
there is {\it not} full cohomogeneity, one must be more disciplined in deriving
a consistent first law, and allow for general variations in the definition of variables.
Note that
\be
\delta J_S = \delta (aM_S) = a \delta M_S - \frac{4\pi}{3} a \ell^2 M_S \delta P\,.
\ee
Thus, if we return to our angular velocity subtraction, and notionally
define\footnote{This form may be motivated by the expression
\eqref{Ominf1}, by identifying $\alpha=K(1-\Xi)/\Xi$, which is a `constant' once
the $a\to \ell$ limit is taken and $K$ becomes a redundant parameter.}
\be
\Omega_\infty = -\frac{\alpha}{a}\,,
\ee
then
\be
(\Omega_S - \Omega_\infty) \, \delta J_S = \Omega_S \,\delta J_S
+ \alpha \delta M_S - \frac{4\pi}{3K} m a^2 \, \frac{\alpha}{1-\Xi}\,,
\ee
that is, our first law \eqref{supFirst} is also consistent (for any fixed $a/\ell$) for the one-parameter family of variables
\be\label{1param}
M_S^\alpha = (1+\alpha) M_S\,, \quad
V_S^\alpha  = \frac{2\mu_S r_+}{3} (r_+^2+a^2)
+\frac{4\pi}{3K} \frac{\alpha ma^2}{1-\Xi}\,,\quad
\Omega_S^\alpha = \frac{a}{r_+^2+\ell^2} + \frac{\alpha}{a} \;.
\ee
Thus there is a one parameter freedom that can be thought of as
the `choice of angular velocity at infinity'. The actual computation yields infinite
$\Omega_\infty$ so one can think of the above value as some sort
of renormalization. One can easily check that for $\alpha<1/2$
the reverse isoperimetric inequality is violated, whereas it becomes
satisfied for $\alpha>1/2$.

Note that similar to what was done in  \cite{Hennigar:2014cfa}, the
thermodynamic quantities \eqref{1param} (in the limit $a\to \ell$)
can be directly obtained from the {\em Superentropic} metric \eqref{supermetric}
by using the conformal method \cite{Ashtekar:1999jx}. Namely, denoting $Q(\xi)$ a conformal
charge corresponding to the Killing vector $\xi$, we find
\be
Q(\partial_t)=M_S\,,\quad Q(\partial_\phi)=J_S\,,
\ee
and therefore
\be
M_S^\alpha = Q(\partial_t-\Omega_\infty \partial_\phi)=(1+\alpha)M_S\,.
\ee

To summarize, simply demanding consistency does not lead to a
unique thermodynamics for the constrained Kerr-AdS black hole,
for which $a = \sqrt{1-\Xi} \ell$. The procedure is ambiguous in that
there is (at least) a 1-parameter family of consistent thermodynamic
quantities some of which do satisfy the isoperimetric inequality and
some of which do not. The origin of this ambiguity is the degeneracy
of the first law, which is no longer of full cohomogeneity and thence
no longer fixes the thermodynamic quantities uniquely. We demonstrated
this explicitly by working down from the full expressions
\eqref{TDcharges}, taking $a = \sqrt{1-\Xi} \ell$, and defining a
set of consistent parameters that can be seen to reduce to the
Hennigar et al.\ expressions in the limit that both $\Xi$ and $\alpha$
tend to zero.

\section{Conclusions}\label{sec:Concl}

In this paper we have discussed the general parameter space for
slowly accelerating black holes and defined a {\it critical limit} in which
at least one of the conical deficits becomes maximal at $\delta = 2\pi$.
We then discussed thermodynamics of these critical black holes, in particular
focussing on the Reverse Isoperimetric Inequality, reviewing a proof of the inequality
and confirming that it holds in the critical limit, which is now smoothly connected
to non-critical black holes. This is manifestly distinct from the argument for
{\it Superentropic Black Holes} \cite{Hennigar:2014cfa} therefore we
have revisited this particular solution and critically examined the arguments
in the literature.

We presented two possible alternate continuous ways of taking the
superspinning limit, one by fixing $a/\ell$ and allowing it to tend to
unity; the second taking an accelerating black hole, fixing
one deficit to its maximal value of $2\pi$ and allowing the other
deficit to approach $2\pi$.
In each case, the fully co-homogeneous thermodynamic parameters
$M$, $V$ and $\Omega$ diverge in the superspinning limit,
thus in order to have finite charges a renormalisation prescription
is required. Using the degeneracy of the thermodynamic variables
that results from fixing $a/\ell$, we showed how the first law can be
reorganised, with a redefinition of the thermodynamic charges that
results in a one parameter family.
This new degree of freedom in turn
raises doubts on the correctness of the superentropic
thermodynamics and gives an alternate argument in favour of
non-superentropic thermodynamics.

It remains to be shown whether similar doubts could arise also for
other {\em Superentropic} black holes, for example the
recently studied charged BTZ black holes \cite{Frassino:2015oca,
Cong:2019bud,Johnson:2019wcq,Johnson:2019mdp} have an 
apparently similar issue. The charged BTZ black hole is a three
dimensional electrically charged solution of the Einstein-Maxwell 
equations \cite{Martinez:1999qi}
\be\label{BTZ}
ds^2 = f(r) dt^2 - \frac{dr^2}{f(r)} - r^2d\varphi^2\,,\quad 
F = \frac{Q}{r} dt \wedge dr\,,
\ee
where $Q$ is the black hole charge, $f(r)$ satisfies
\be
f' = \frac{2r}{\ell^2} - \frac{8\pi GQ^2}{r} \,,
\label{eqforf}
\ee
and $\ell$ is the AdS radius, defined by $\Lambda = -1/\ell^2$.
In \cite{Martinez:1999qi}, the Newton's constant is fixed by setting
$16\pi G = 1$, however, we will temporarily retain
this dimensionful parameter, writing $8\pi G = L_p=M_p^{-1}$ the 
3D Planck length, in order to emphasise the source of
superentropicity. 

Integrating \eqref{eqforf} yields the potential
\be
f(r) = \frac{r^2}{\ell^2} - L_p Q^2 \log \left ( \frac{r}{r_0} \right) - 2m\,,
\ee
where $m$ is an integration constant we identify as the mass parameter, 
and $r_0$ is a dimensionful integration parameter inserted to render
the argument of log dimensionless. This is in part the reason for maintaining the
dimensionful parameter $G$, if we set $8\pi G=1$, or $1/2$ as in the original
paper \cite{Martinez:1999qi}, then 
$r$ becomes dimensionless, and we need not add any scale
inside the logarithm, however, in keeping with convention, we introduce
the scale $r_0$ as a second integration constant.

The extended thermodynamics of this black hole has been studied 
in \cite{Gunasekaran:2012dq,Frassino:2015oca}, and is shown to 
crucially depend on the choice of the integration parameter $r_0$.
The conventional choice in the literature is to set $r_0=\ell$, however, in extended
thermodynamics this has an important consequence: $\ell$ is related
to the thermodynamic pressure, so varying $P$ has the consequence
of varying the integration constant. Imposing this value of $r_0$ leads to
the thermodynamic variables
\be\label{TDSposs1}
\beal
M&= \frac{m}{4}=\frac{r_+^2}{8\ell^2}
-\frac{Q^2}{16}\log\bigl(\frac{r_+}{\ell}\bigr)\,,&
T& =\frac{r_+}{2\pi \ell^2}-\frac{Q^2}{8\pi r_+}\,,&
S& =\frac{\pi r_+}{2}\,, \\
V & =\pi r_+^2-\frac{1}{4}Q^2\pi \ell^2\,,&
\Phi&=-\frac{1}{8}Q\log\bigl(\frac{r_+}{\ell}\bigr)\,,&
P &=\frac{1}{8\pi \ell^2}\,,
\eeal
\ee
that obey the standard 1st law and Smarr relations:
\be\label{BTZ1stSmarr}
\delta M=T\delta S+\Phi \delta Q+V\delta P\,,\quad TS=2PV\,,
\ee
where we have set $L_p=1/2$ to align with the literature \cite{Martinez:1999qi}.

Note the non-geometric correction to the black hole volume $V$, 
originating from the aforementioned variation of the integration constant.
This is precisely the term that implies the violation of the
reverse isoperimetric inequality. Although preferred in \cite{Frassino:2015oca}, 
this option is questionable from various perspectives. First, the potential 
$\Phi$ in \eqref{TDSposs1} is that of the electrostatic potential evaluated 
on the horizon, however this is not gauge invariant; usually one takes the
potential difference between the horizon and infinity as a gauge invariant
thermodynamic potential, however this is problematic in 3D as the potential 
at infinity obviously diverges. Secondly, the introduction of $r_0$ can be viewed
as part of a renormalisation procedure, and indeed is discussed as such in 
\cite{Martinez:1999qi}. If a cutoff is introduced, then one would expect
that this cutoff would remain fixed as one is varying {\it physical} parameters
in a thermodynamic process. This perspective leads to an alternative
formulation of thermodynamics, where we identify $r_0$ in \eqref{BTZ} 
as `enclosing' the BTZ black hole in a circle of radius $r_0$ as in 
\cite{Martinez:1999qi,Cadoni:2007ck}. Upon this, the potential at 
`infinity' ($r=r_0$) vanishes and we obtain the following thermodynamic 
quantities \cite{Frassino:2015oca}:
\be\label{TDSposs2}\beal
M&= \frac{m}{4}=\frac{r_+^2}{8\ell^2}
-\frac{Q^2}{16}\log\bigl(\frac{r_+}{r_0}\bigr)\,,&
T& =\frac{r_+}{2\pi \ell^2}-\frac{Q^2}{8\pi r_+}\,,&
S& =\frac{\pi r_+}{2}\,, \\
\Phi&=-\frac{1}{8}Q\log\bigl(\frac{r_+}{r_0}\bigr)\,,&
V & =\pi r_+^2\,,&
P &=\frac{1}{8\pi \ell^2}\,,
\eeal\ee
together with the circumference $C=2\pi r_0$ and a dual thermodynamic 
potential $\mu_C = Q^2/16C$ if one wishes to vary the physical enclosure
around the black hole. With these the first law \eqref{BTZ1stSmarr} remains 
satisfied, but the Smarr relation now picks up a $C\mu_C$ term due
to the scaling properties of $r_0$. The thermodynamics \eqref{TDSposs2} 
is obviously non-superentropic. 

We therefore suspect that the traditional thermodynamics of this 
somewhat pathological solution is most likely not the correct one.
Finally, in an interesting recent twist, the rotating (uncharged) BTZ black 
hole, which as a traditional Einstein solution is not superentropic, can be
reimagined as a solution of the gravitational Chern--Simons action 
\cite{Frassino:2019fgr, Cong:2019bud}. The thermodynamic parameters 
become ``exotic'' (with mass and angular momentum charges reversed and 
the entropy no longer given by the horizon area). This it seems may also 
violate the reverse isoperimetric inequality, however since the {\it Reverse 
Isoperimetric Inequality} conjecture of \cite{Cvetic:2010jb} was originally
put forward for Einstein gravity, this can not be regarded as a counter-example.
What it abundantly clear however is that the existence and origin of 
superentropicity most certainly deserves further investigation.

\acknowledgments

MA was supported by an STFC studentship, and LC by CONACyT.
RG is supported in part by the STFC [consolidated grant ST/P000371/1],
and in part by the Perimeter Institute.
PK acknowledges the Czech Science Foundation Grant 19-01850S;
the work was done under the auspices of the Albert Einstein
Center for Gravitation and Astrophysics, Czech Republic.
DK is also supported in part by Perimeter, and by the NSERC.
PK, MA, and LC would also like to thank Perimeter Institute for
hospitality while this research was undertaken.
Research at Perimeter Institute is supported in part by the Government
of Canada through the Department of Innovation, Science and Economic
Development Canada and by the Province of Ontario through the Ministry
of Economic Development, Job Creation and Trade.
RG would also like to thank the Aspen Center for Physics for hospitality while some
of this work was being undertaken.
Work at Aspen is supported in part by the National Science
Foundation under Grant No. PHYS-1066293.

\end{document}